\documentclass[conference]{IEEEtran}

\usepackage{balance}
\usepackage{float}
\usepackage{caption}
\usepackage{subcaption}
\usepackage{tikz,pgfplots,pgfplotstable,booktabs}
\usepackage[official]{eurosym}
\usepackage{xcolor}
\usepackage{colortbl}
\usepackage{comment}
\usepackage{cite}
\usepackage{circuitikz}
\usepackage{amsmath,amsfonts,amsthm,bm}

\newcommand\blfootnote[1]{%
  \begingroup
  \renewcommand\thefootnote{}\footnote{#1}%
  \addtocounter{footnote}{-1}%
  \endgroup
}

\usepackage{hyperref}
\usepackage{tikz}

            \usepackage{placeins}
\newcommand{\subparagraph}{}
\usepackage{titlesec}

\titlespacing*{\section}{0pt}{3pt}{0pt}
\titlespacing*{\subsection}{0pt}{2pt}{0pt}
\titlespacing*{\subsubsection}{0pt}{0pt}{0pt}
\newcommand{\norm}[1]{\left\lVert#1\right\rVert}

\allowdisplaybreaks[1]
\usepgfplotslibrary{fillbetween}
\usepgfplotslibrary{statistics}
\usepackage{hyperref}

\pgfplotsset{
    discard if not/.style 2 args={
        x filter/.code={
            \edef\tempa{\thisrow{#1}}
            \edef\tempb{#2}
            \ifx\tempa\tempb
            \else
                
            \fi
    }},
    boxplot/hide outliers/.code={
        \def\pgfplotsplothandlerboxplot@outlier{}}
    }

\title{Outing Power Outages: Real-time and Predictive Socio-demographic Analytics for New York City}

\author{Samuel Eckstrom, Graham Murphy, Eileen Ye, Samrat Acharya, Robert Mieth and Yury Dvorkin}

\thanks{Test}

\begin{document}
\bstctlcite{IEEE:BSTcontrol} 

\maketitle

\begin{abstract}
Electrical outages continue to occur despite technological innovations and improvements to electric power distribution infrastructure. In this paper, we describe a tool that was designed to acquire and collect data on electric power outages in New York City since July 2020. The electrical outages are then displayed on a front-end application, which is publicly available. We use the collected outage data to analyze these outages and their socio-economic impacts on electricity vulnerable population groups. We determined that there was a slightly negative linear relationship between income and number of outages. Finally, a Markov Influence Graph was created to better understand the spatial and temporal relationships between outages.
\end{abstract}

\section{Introduction}

\blfootnote{This work was supported in part by the National Science Foundation (NSF) under Award EECS-1847285 and EECS-2029158.}

Most aspects of our modern society require reliable access to electric power.
Therefore, it is imperative that the electric grid is offline as little as possible to avoid the shortfall of the technologies that enable our everyday life.
However, achieving 100\% reliability is often impossible or prohibitively expensive to most private and commercial electricity consumers. 
Despite the best efforts of power utility companies, service outages caused by equipment failures or external conditions such as damage from fire, storm or vegetation remain  common. 
While such outages are disruptive to all private and commercial consumers, some population groups are especially vulnerable to electricity service disruption and should be prioritized in the communication and restoration efforts of the utility. 
This paper presents an interactive decision-support tool that allows to evaluate outages in their socio-demographic context, i.e. not only outage parameters are evaluated but also their impacts on population, and helps to mitigate and repair future outages. 

\subsection{Background and motivation}
The impact of electrical outages on everyday life cannot be understated as they severely disrupt both the health and economics of a population. 
Electrical outages can have both direct and indirect economic effects on a population. While direct cost of outages, such as the cost to repair or replace the defective infrastructure, are often more easily determined and, thus, more perceivable, short- and long-term indirect costs can cause serious economic inefficiencies. Researchers estimate 5 to 75 billion dollars in yearly direct cost of electrical outages over the entire US. 
The indirect cost over the US are much more difficult to predict, but it appears that the direct and indirect costs of electrical outages decrease the US economic output by roughly 1\% \cite{sanstad2020case}. 
Due to the incredibly interconnected nature of economics,  such indirect economic losses can be quite significant, e.g., by cascading disruptions of supply chains \cite{schmidthaler2016assessing, cameron2012network}.

In addition to the economic effects of electrical outages, the shortfall of power supply also severely affects population health. 
For example, the risk of carbon monoxide poisoning, due to failing alarm systems, and temperature related illnesses such as heat stroke or hypothermia, due to failing temperature regulation systems, significantly increases during outages. 
Further, people depending on electric medical devices will be severely affected~\cite{casey2020power}.

The increased frequency of extreme weather events, e.g., heatwaves and storms, amplify both the probability and the severity of electricity outages. 
For example, New York City's utility \textit{Consolidated Edision} (``Con Edison'') frequently asks its customers to save power during high-heat weather as the additonal load from residential air conditioners threatens system reliability \cite{coned_heatwave}. 
While outages are more likely during extreme weather events, they are also more impactful especially to those parts of the population that are more susceptible to temperature changes.
These groups include the elderly; people with pre-existing conditions, such as: respiratory, mental health, and cardiovascular issues; people taking medicine that affects their natural body-temperature regulation; people with either cognitive or physical disabilities that bar them from reacting quick enough to sudden temperature changes; people living in hazardous areas; and people that do not have the economic assets to protect themselves. 
Vulnerabilities in the latter group may be caused by insufficient investments into cooling and heating system maintenance or building insulation, which cause a more rapid and severe change of ambient temperature \cite{klinger2014power}.

The relationship between the occurrence and impact of power outages for certain demographic groups can be analyzed in terms of energy justice or \textit{electricity vulnerability}. See \cite{jessel2019energy} for a comprehensive review, where electricity vulnerability is discussed in terms of various socio-economic dimensions.
While these concepts have been studied and highlighted in the context of \textit{ex post} analysis, see e.g., \cite{hernandez2016understanding}, real-time and predictive data analytics that capture electricity vulnerability are lacking. 
However, such time-sensitive outage analyses become more important with the emergence of novel resilience solutions, such as mobile battery systems, micro-grids with islanding capabilities and cyberphysical smart grids \cite{kim2018enhancing, wang2015research, savena2017cost}.
To address this need, this paper describes an interactive, publicly accessible dashboard (\url{outagesnyc.hosting.nyu.edu}) to understand the spatial and temporal occurrence of electrical outages in New York City and provides a statistical analysis of  their impact on different socio-economic groups. 
To this end we combine real-time outage data with a comprehensive demographic information for every zip code in New York City. 
The resulting tool consists of three components: (i) a dashboard, which visualizes  outages are occurring in real time, (ii) statistical analyses of historical outages, and (ii) analytics tool for predicting future outages.

\subsection{Overview}

The developed tool crawls, analyzes and displays real-time outages obtained from the Con Edison Outage Map \cite{coned_outagemap}. {\color{black}
Unlike \cite{coned_outagemap}, which only shows real-time outages, the developed dashboard stores the captured data to create a comprehensive archive of outages, which is publicly available to facilitate further research. 
}
On an interactive map, it shows current, historical and predicted outages and a color-coded electricity vulnerability ranking, called \textit{electricity vulnerability index}.
Section~\ref{sec:dashboard} provides details on the development of the dashboard and the electricity vulnerability index.
Section~\ref{sec:statistical_analysis} examines outage occurrence and correlations to household income and racial demographics. 
Exploring the collected outage data, we found, e.g., that a disproportionate amount of outages occurred in zip codes with family median incomes under \$125,000. 
Next, to address outages more effectively and to better understand the spatial dependencies between outages, we created a Markov influence graph that supports the prediction of outages in predefined areas in the city. Section~\ref{sec:influence_graph} presents this work. 
Finally, Section~\ref{sec:conclusion_and_future_work} provides a concluding discussion of this work and outlines future research.

\section{Dashboard and Outage Ranking}
\label{sec:dashboard}

\begin{figure}
    \centering
    \includegraphics[width=0.85\linewidth]{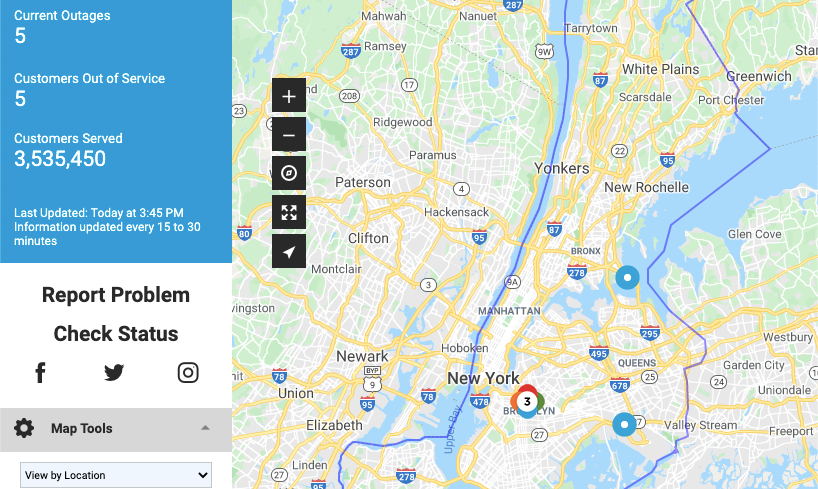}
    \caption{Con Edison Outage Map. As Can be seen in the map, there are a total of five outages at this specific moment. There are three symbols currently on the map. Two of the symbols are small blue circles with a smaller white circle inside. These symbols represent a single outage. In addition, there is another symbol that represents a cluster of outages. In this case, this symbol has a multicolored border with a small number '3' on the inside. Therefore, we know that if we zoom in closer, there will be three outages at this location. }
    \label{fig:coned}
\end{figure}

The central objective of this work is to obtain real-time and historical outage data, put them into their demographic context via a suitable metric and visualize them in an accessible manner.
This section describes the used data processing and visualization methods.
The resulting dashboard can be accessed at \textbf{\url{outagesnyc.hosting.nyu.edu}}.

\subsection{Online Dashboard}

The primary data source of our dashboard is the Con Edison Outage Map \cite{coned_outagemap}, shown in Fig.~\ref{fig:coned}.
To collect real-time outage data from this resource, we created a \textit{Selenium} web crawler using \textit{Python} 3.6 to extract the physical addresses of the outages placed in the Con Edison Outage Map.
This crawler runs continuously in multiple simultaneous instances on a virtual machine set up on \textit{Google Cloud Platform}. 
When the information for each outage is obtained, we send the address and the current time stamp to a \textit{Microsoft Azure} database. 
This database stores each outage in one of three tables: ``Currently crawled'' (but not yet processed), ``processed'', and ``historical'' (when it is removed from the Con Ed outage map). 
The database is updated every thirty minutes by a \textit{Python} script that monitors the status of each outage. 
The first table is the initial entry point for every crawled outage and consists of exactly those outages that are currently displayed in the Con Ed outage map.
The \textit{Python} script then takes the crawled outages and inserts them into the table for processed outages. 
Next, if an outage exists in the table for processed outages, but not anymore in the table for currently crawled outages, it is inserted into the historical outage table and removed from the processed outage table.
When an outage is processed, the \textit{Python} script automatically obtains additional information on this outage, i.e.zip code and name of the New York City borough\footnote{Administrative divisions of New York City, i.e., The Bronx, Brooklyn, Manhattan, Queens, Staten Island.} via the \textit{Google Maps API}.
To this end, we use a \textit{cron} job, a job scheduler for Unix-like operating systems, to execute \textit{JavaScript} code that communicates with the \textit{Google Maps API}.

\begin{figure}
    \centering
    \includegraphics[width=0.8\linewidth]{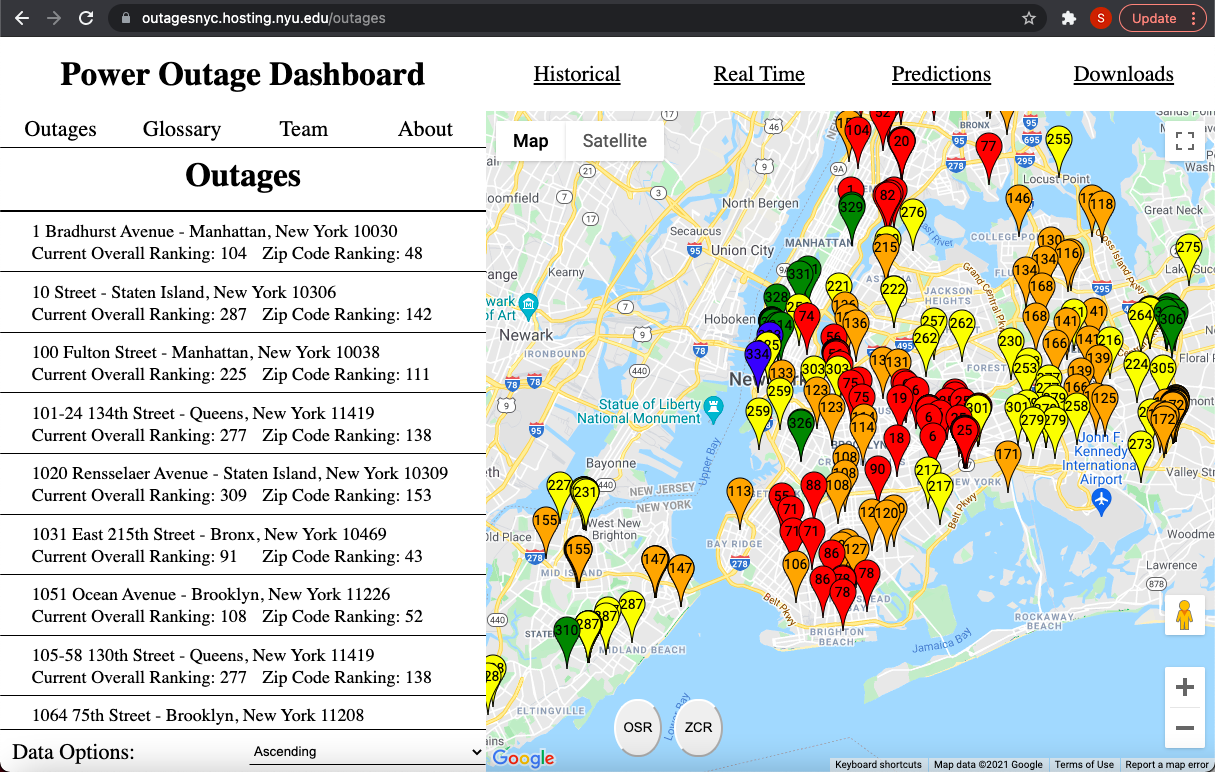}
    \caption{The real time outages page of our dashboard. As can be seen, many different outages have occurred with different electricity vulnerability severity. }
    \label{fig:dashboard}
\end{figure}

To visualize the data, we created a website using a \textit{React} front end and an \textit{Express} back end. 
The website has four pages: ``Historical'', ``Real Time'', ``Predictions'', and ``Downloads''. 
The website initially directs to the real-time outages page, as shown in Fig.~\ref{fig:dashboard}. 
On this page, the current outages are placed on an interactive map (provided by \textit{Google Maps}) and color-coded based on their electricity vulnerability index (see Section~\ref{ssec:electricity_vul_index} below for details). 
Further, this page provides four additional tabs: ``Outages'', ``Glossary'', ``Team'', and ``About''. 
The outages tab itemizes all outages shown on the map in a continuous list. 
The glossary tab defines important terminology to guide the user.
The team tab gives a brief description of all collaborators of this project.
Finally, the about tab provides information on the motivation and background of the project. 

The page labeled ``Historical'' contains a histogram of all the outages that we have scraped and a map with all the demographic information that we have collected. (See Section~\ref{sec:statistical_analysis} below for a more detailed description of the demographic data).  
The page labeled ``Predictions'' is currently under construction and will be available at a later date. This page will consist of a visualization for the Markov influence graph (see Section~\ref{sec:influence_graph}) and it's practical applications, e.g., for outage prediction and optimal repair crew routing.
Finally, the page labeled ``Downloads'' allows the user to download the data that we have collected, which gives them the opportunity to perform their own analyses and verify ours. 
Each link on this page will download the data that is associated with one of the tables described above.

\subsection{Electricity Vulnerability Index}
\label{ssec:electricity_vul_index}

The developed electricity vulnerability is an ordinal ranking that uses various socio-demographic features. 
First, \textcolor{black}{we identified the relevant socio-demographic features  from \cite{klinger2014power,jessel2019energy} as}: percent elderly (i.e., people older than 65), number of cooling centers, number of affordable housing buildings, number of affordable housing units, percent below poverty, number of children under the age of five, and the average number of caregivers. 
For each zip code, we obtained data on these features from the U.S. Census Bureau \cite{uscencus} and the New York City Open Data platform \cite{nycopendata}; normalized each feature by subtracting the minimum value over all zip codes, then dividing by the difference of the maximum and minimum values over all the zip codes; and computing the equally weighted sum of all feature values for each zip code. 
{\color{black}While it is possible to assign feature weights, it is difficult to determine exact values given the repercussions to public morbidity and mortality. A possible approach would assign weights to minimize loss of Quality-Adjusted Life Years (QALY). However, determining the QALY and required feature weights for each zip code is complex and beyond the scope of this work. 
We therefore opt for an equally weighted feature sum as a transparent method.
} 
Sorting these sums creates the ordinal ranking that is our electricity vulnerability index for each zip code. 

On our dashboard, we display two different versions of this index: The overall severity ranking (OSR) and the zip code ranking (ZCR). 
The overall severity ranking is the severity of the outage relative to all the other current outages.
The zip code ranking gives the ranking relative to all zip codes in New York City. 
Therefore, the overall severity ranking can be anything from 1 to $N$, where $N$ is the number of current outages, and the zip code ranking can be anything from 1 to 241, where 241 is the number of zip codes in New York City. 
The ranking for each zip code is displayed on the Google Maps markers and the outage list in the left tab. Further, the ranking is indicated by the color of the marker, which is chosen as:
\begin{itemize}
    \item[\tikz{\draw[fill=red,line width=0.5pt]  circle(0.8ex);}] 
    Outages occurring in zip codes that are ranked as very vulnerable (ranked 1--50) are displayed in red.
    \item[\tikz{\draw[fill=orange,line width=0.5pt]  circle(0.8ex);}]
    Vulnerability rankings 51--100 are shown in orange.
    \item[\tikz{\draw[fill=yellow,line width=0.5pt]  circle(0.8ex);}]
    Vulnerability rankings 101--150 are shown in yellow.
    \item[\tikz{\draw[fill=green!50!black,line width=0.5pt]  circle(0.8ex);}]
    Vulnerability rankings 151--200 are shown in green.
    \item[\tikz{\draw[fill=blue,line width=0.5pt]  circle(0.8ex);}]
    Vulnerability rankings 201--$\infty$ are shown in blue.
\end{itemize}

\section{Statistical Analysis of Outage Data}
\label{sec:statistical_analysis}

Building on the collected outage data from July 2020 to September 2021 (as described in Section~\ref{ssec:electricity_vul_index}), we performed statistical analyses to relate the frequency and location of electricity outages in New York City to the soci-demographic features of their respective zip codes.

\subsection{Socio-Demographic Analysis}

\begin{figure}
    \centering
    \includegraphics[width=0.95\linewidth]{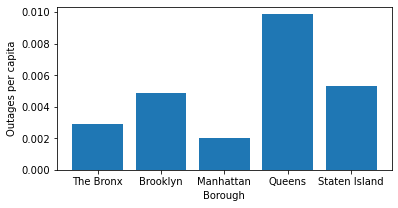}
    \caption{Outages by borough per capita}
    \label{fig:outages_per_capita}
\end{figure}

First, we consider the total number of outages per borough. 
The results, adjusted by the population of each borough, are reported in Fig.~\ref{fig:outages_per_capita}.
Brooklyn and Queens recorded the highest absolute number of reported outages, with the total of over 12,360 outages in Brooklyn and 15,200 outages in Queens. 
Meanwhile, The Bronx, Manhattan and Staten Island only experienced 3,281, 3,180, and 2,514 outages, respectively.

\begin{figure}
    \centering
    \includegraphics[width=0.95\linewidth]{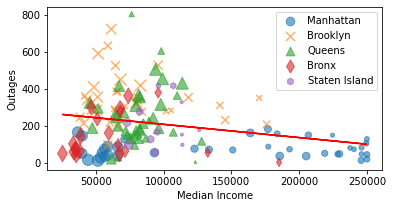}
    \caption{Outage occurrences in a zip code relative to its median family income; Different markers indicate boroughs and marker size corresponds to percentage of non-white citizens (larger=higher); Trend line for all data points.\protect\footnotemark }
    \label{fig:income_v_outages}
\end{figure}

\begin{figure}
    \centering
    \includegraphics[width=0.95\linewidth]{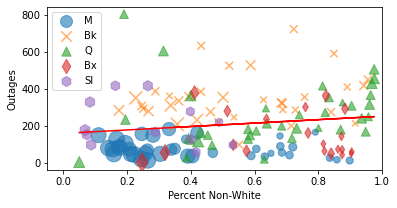}
    \caption{Outage occurrences in a zip code relative to its non-white population ratio; Different markers indicate boroughs and marker size corresponds to median household income (larger=higher); Trend line for all data points.\protect\footnotemark[3]}
    \label{fig:race_outages}
\end{figure}

Next, we analyzed the outage data by examining the number of outages per zip code.
Additionally, we paired the zip-code outage data with various demographic features, which we obtained from the U.S. census bureau \cite{uscencus}.
As example Fig.~\ref{fig:income_v_outages} shows the relationship between the number of outages and  the median income of all New York City zip codes, and Fig~\ref{fig:race_outages} shows the number of outages in a zip code relative to the percentage of non-white citizens at that zip-code.
In Fig.~\ref{fig:income_v_outages}, we observe that zip codes with high outage counts are also zip codes with relatively low median incomes.
The data points indicate a negative linear relationship between family median income per zip code and the number of outages.
Moreover, the data in Fig.~\ref{fig:income_v_outages} shows significantly fewer outages in zip codes with  median family incomes above \$125,000.
In Fig~\ref{fig:race_outages}, on the other hand, we observe a very slight positive relationship between the outage numbers in a zip code and the percentage of non-white population.

Further, Figs.~\ref{fig:income_v_outages} and \ref{fig:race_outages} reveal significant differences between boroughs.  
Queens' and Brooklyn's infrastructure is characterized by larger numbers of  single-family houses and above-ground distribution system infrastructure, which makes these boroughs more susceptible to weather-related outages. 
At the same time, zip-codes with these infrastructure characteristics exhibit a relatively lower median income and a relatively higher non-white population.
Similarly, The Bronx, while mainly consisting of zip codes with a relatively low median income, shares infrastructure characteristics with Manhattan, i.e., multi-family apartment buildings and below-ground distribution system infrastructure.
Given the characteristics of New York city, additional spatial analyses of infrastructure and outage characteristics, as well as population density would be insightful. 
As this data is not available from our current data set, we will postpone such analyses to future studies.


\footnotetext{\label{fn:subtrends}\color{black} Note that trends within boroughs take individually may deviate from this city-wide trend line.}

\subsection{Outage Causes}

\begin{figure}
    \centering
    \includegraphics[width=0.95\linewidth]{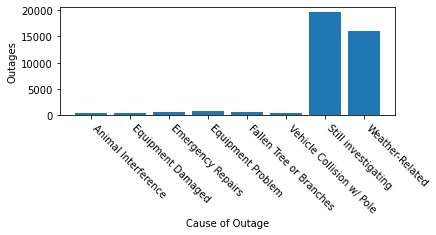}
    \caption{Outage occurrence by cause.}
    \label{fig:outage_cause}
\end{figure}

Further, we explored the most common causes of outages reported by Con Edison.
First, as shown in Fig.~\ref{fig:outage_cause}, outage causes are labeled as being investigated.
This is reasonable, because we are obtaining outage data in real-time, i.e., potentially before a clear cause has been identified.
Also, investigations could be delayed if Con Edison experience large numbers of outage reports.
Most outages with known causes are weather related. 
Other reported outage causes include animal interference, damaged our faulty equipment, emergency repairs, damage from fallen trees or branches, and vehicle collisions with poles. However, these causes are clearly dominated by weather-related causes.

\section{Influence Graph for Outage Prediction} 
\label{sec:influence_graph}

\begin{figure}
    \centering
    \includegraphics[width=0.95\linewidth]{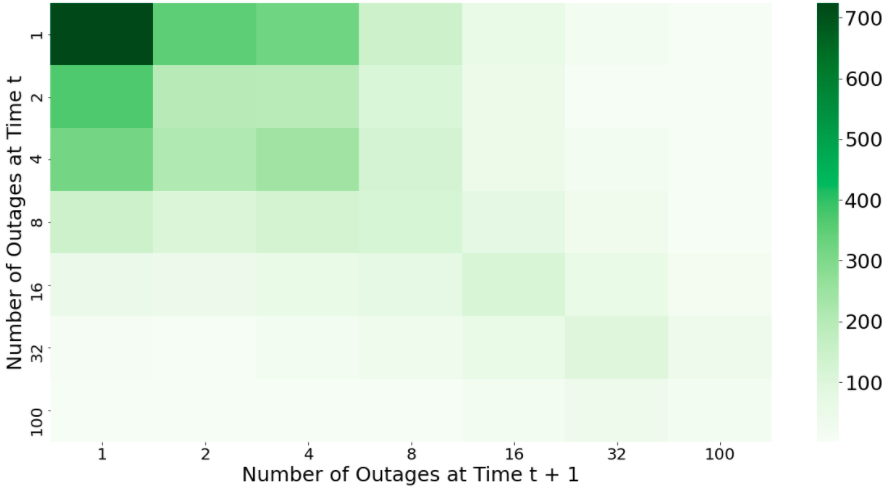}
    \caption{Correlation matrix depicting how varying numbers of outages cascade throughout time. X- and Y-axis show the upper limit of outages counted into the respective bin.}
    \label{fig:spatialcorrelationmatrix}
\end{figure}

This section describes data-driven analyses that can support the predictions of future outages.
First, we analyze the inter-temporal behavior of outage numbers to verify the existence of any predictability within the collected data.
For this we grouped the collected outage data by time stamp and zip code. 
Next, we created a two-dimensional matrix that captures the correlation between outage occurrences as follows: (i)~Determining the total number of outages that occurred over all zip codes for each time step. Note that the length of a time step can be freely chosen to tune the desired temporal resolution. In our analyses we found a two-hour period for each time step to work best.
(ii) Defining seven discrete bins to categorize the number of outages at each time step. The bins are: 0--1, 2, 3--4, 5--8, 9--16, 17--32, and 33--$\infty$ outages. 
(iii) Counting how often the total number of outages transitioned from one bin to another bin over all observed time steps. 
Note that if the bin does not change in between time steps, this is counted as a transition from a bin to itself.
The resulting correlation matrix can be seen in Fig.~\ref{fig:spatialcorrelationmatrix}.

Next, to be able to predict the changes of outage numbers over time, we created a \textit{Markov influence graph} \cite{zhou2020markovian}. 
To this end, we first grouped all outages into 11 zip code sections using k-means (implemented as in \cite{zaki2020data}).  
This method with 11 clusters reliably created clusters with roughly an equal number of outages at each time step. 
To apply this method, we first assign a single latitude and longitude to each zip code by identifying its centroid. 
Next, we initialized the k-means algorithm by randomly choosing 11 initial zip codes and setting them as the central cluster coordinates.
Then, we sequentially inserted each remaining zip code into the 11 clusters, such that each new zip code is added to the cluster whose centroid is closest to the centroid of each new zip code. After the addition of each zip code, the coordinates of the cluster centroid are recalculated. 
The resulting 11 clusters contain all zip codes such that their physical distance within the clusters is minimized.
Fig.~\ref{fig:influencegraph} shows the resulting clusters on a map of New York City.

\begin{figure}
    \centering
    \includegraphics[width=0.95\linewidth]{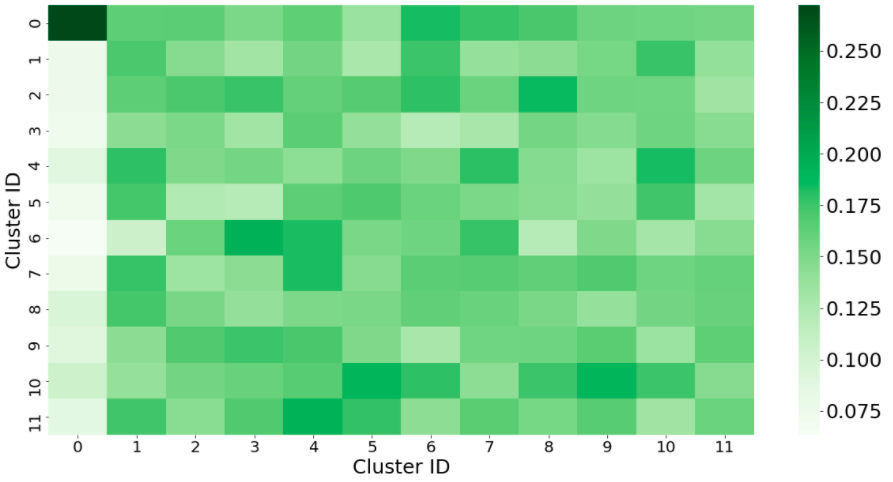}
    \caption{Markov transition matrix as basis for influence graph.}
    \label{fig:transitionmatrix}
\end{figure}

\begin{figure}
    \centering
    \includegraphics[width=0.9\linewidth]{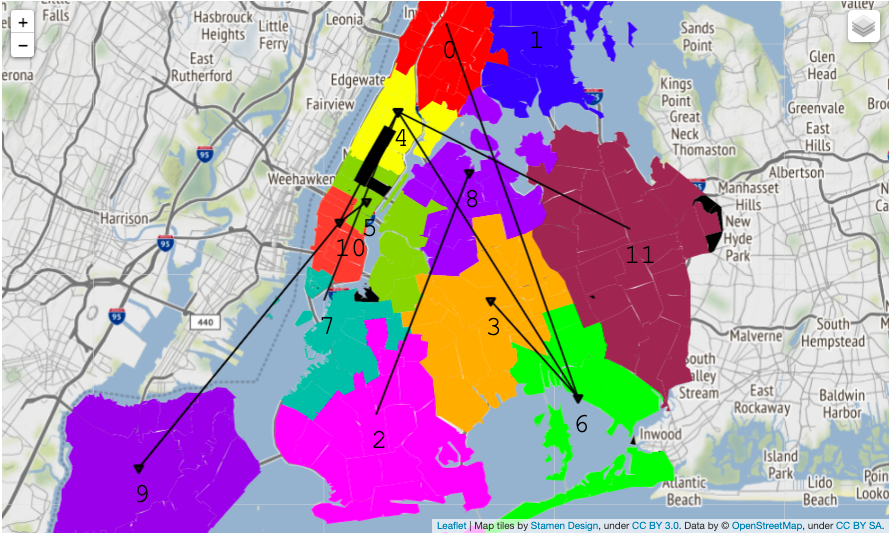}
    \caption{Influence graph over New York City showing the spatio-temporal relationships between zip code clusters.
    shown via arrows.
    The graph is shown via arrows that represent the 10 largest weights of the transition matrix from Fig.~\ref{fig:transitionmatrix}.
    The start of each arrow is associate with the state at a time $t$ and the end with the state at $t + 1$.
    }
    \label{fig:influencegraph}
\end{figure}

Using the computed zip code clusters, we group all outages into their respective clusters via their zip code information. 
Then, we compute the number of outages in each cluster for each time step. 
Recall, that the desired time resolution, i.e., the amount of time for each time step, can be freely chosen. As for the correlation matrix above, we use two-hour time steps. 
For each time step, we can now define two $11\times 1$ column vectors $o$ of outage numbers: one for the time step $t$ itself and one for the next time step $t + 1$. 
We use these vector pairs to create a transition matrix that can predict the number and location of outages at a time $t+1$ from observing the number and location of outages at time $t$.
We obtain such a transition matrix efficiently by using repeated random sampling. For each time step, we created $s=\{1,...,100000\}$ random candidate matrices $T^s_t$ and identified the matrix $T^{*}_t$ that minimizes the mean square error between the true outage vector $o_{t+1}$ and predicted outage vector $\hat{o}_{t+1}^s = T^s_t o_t$, i.e., $T^{*}_t = \arg \min_{\{T^s_t,\ \forall s\}} \norm{o_{t+1} - T^s_t o_t}^2_2\ \forall t$. 
We obtain the final transition matrix $T$ as the average of all $\{T^{*}_t,\ \forall t\}$.
The resulting transition matrix can be seen in Fig.~\ref{fig:transitionmatrix}.
 
{\color{black}We can use this transition matrix to make inferences on future outages. To do so, we multiply the transition matrix with a new outage vector $o_{\text{new}}$ and obtain the expected number of outages for each location in the next time step. 
A comprehensive presentation and validation of this method is beyond the scope of this paper and will be presented in an upcoming publication.
}
The average transition weights between each cluster are the resulting influence graph. 
This graph shown in Fig.~\ref{fig:influencegraph} plotted over a map of New York City.
Note that Fig.~\ref{fig:influencegraph} shows the 10 most influential weights to avoid clutter. 

\section{Conclusion} 
\label{sec:conclusion_and_future_work}

This paper presented an interactive online tool to acquire, analyze and visualize real-time data on electric power outages in New York City. 
We have explained the technical details of the tool development and proposed a electricity vulnerability index to evaluate the socio-demographic impact of these outages. 
Further, we conducted additonal statistical analyses on the relationship between outage numbers and various demographic features. 
Finally, this paper demonstrated a method to analyze the spatio-temporal relation between outages and derived a Markovian influence graph that supports outage prediction.

We believe that the methods and socio-demographic analyses shown in this paper are of interest to utility companies, researchers, and the public. 
In a next work-package, we will obtain feedback from Con Edison on our work to align our methods with the needs of a real-world utility. 
{\color{black}For example, our socio-demographic ranking and predictive analytics can be combined with algorithms that determine the planning and real-time routing of repair crews and effective communication to affected households.}
{\color{black}Further, we will explore enhancements of the predictive analytics using additonal data sets, e.g., weather forecasts.}
Similarly, such analyses provide suitable means to evaluate the social impact of infrastructure investments and, ultimately, support the development of a more just power system.

\bibliographystyle{IEEEtran}
\bibliography{literature}

\end{document}